\documentclass[11pt]{article}

\usepackage{graphicx}
\usepackage{amsmath}
\usepackage{amsfonts}
\usepackage{subfigure}
\usepackage{multirow}
\usepackage{amssymb}
\usepackage{graphicx}
\usepackage{algorithmic}
\usepackage{lscape}
\usepackage{longtable}
\usepackage{theorem}
\usepackage{booktabs}
\usepackage{listings}
\usepackage{color}
\usepackage{float}
\usepackage{siunitx}

\usepackage[english]{babel}
\usepackage[round]{natbib}
\usepackage[latin1]{inputenc}
\usepackage{times}
\usepackage[T1]{fontenc}

\title{Modeling Dependence Dynamics of Air Pollution: Pollution Risk Simulation and Prediction of PM$_{2.5}$ Levels}

\begin{document}

\maketitle

\begin{center}

\large Halis Sak \footnote[1]{Corresponding author. Tel: +86.512.88161000-4886 \\ 
\textit{Email addresses:} halis.sak@gmail.com (Halis Sak), guanyu.yang@yahoo.com (Guanyu Yang), bailiang.li@xjtlu.edu.cn (Bailiang Li), wfli@hku.hk (Weifeng Li)}, Guanyu Yang\\ \normalsize
Department of Mathematical Sciences, Xi'an Jiaotong-Liverpool University, Suzhou, China
\\[6pt]

\large Bailiang Li\\ \normalsize
Department of Environmental Science, Xi'an Jiaotong-Liverpool University, Suzhou, China\\[6pt]

\large Weifeng Li\\ \normalsize
Department of Urban Planning and Design, The University of Hong Kong, Hong Kong\\[6pt]

\end{center}

\begin{abstract}

The first part of this paper introduces a portfolio approach for quantifying the risk measures of pollution risk in the presence of dependence of PM$_{2.5}$ concentration of cities. The model is based on a copula dependence structure. For assessing model parameters, we analyze a limited data set of PM$_{2.5}$ levels of Beijing, Tianjin, Chengde, Hengshui, and Xingtai. This process reveals a better fit for the t-copula dependence structure with generalized hyperbolic marginal distributions for the PM$_{2.5}$ log-ratios of the cities. Furthermore, we 
show how to efficiently simulate risk measures clean-air-at-risk and conditional clean-air-at-risk using importance sampling and stratified importance sampling. Our numerical results show that clean-air-at-risk at 0.01 probability level reaches up to $352 \,\SI{}{\micro\gram\meter^{-3}}$ (initial PM$_{2.5}$ concentrations of cities are assumed to be $100 \,\SI{}{\micro\gram\meter^{-3}}$) for the constructed sample portfolio, and proposed methods are much more efficient than a naive simulation for computing the exceeding probabilities and conditional excesses.

In the second part, we predict PM$_{2.5}$ levels of the next three-hour period of four Chinese cities, Beijing, Chengde, Xingtai, and Zhangjiakou. For this purpose, we use the pollution and weather data collected from the stations located in these four cities. Instead of coding the machine learning algorithms, we employ a state-of-the-art machine learning library, Torch7. This allows us to try out the state-of-the-art machine learning methods like long short-term memory (LSTM). Unfortunately, due to small data size and lots of missing values (when we combined the features of the cities) in the data, LSTM does not perform better than a multilayer perceptron.  However, we get a classification accuracy above 0.72 on the test data. 

\end{abstract}

\section{Introduction}

Air pollution has been an increasing issue in many countries. Among various types of air pollutants, particulate matter (PM) has been widely considered as a contributing factor that may significantly affect human health.  
Atmospheric PM mainly comes from six sources: traffic, industry, fuel burning, natural sources (e.g., dust and sea salt) and secondary particles from chemical reactions of primary gaseous pollutants (e.g., NO$_2$, NH$_3$, SO$_2$, and non-methane volatile organic compounds NMVOCs \citep[see e.g., review paper by][]{Karagulianetal;2015}.  Many studies have indicated that particulate air pollution may increase the risk of heart attacks, aggravate asthma, decrease lung function, and even cause lung cancer \citep[][]{Atkinsonetal:2010,Cadelisetal:2014,Correiaetal:2013,Fangetal:2013,Meisteretal:2012}. Particularly, the exposure to particulate matter smaller than 2.5 \SI{}{\micro\meter}, or PM$_{2.5}$, has been found more dangerous and associated with increasing cardiovascular, and respiratory mortality in many cities \citep[][]{Shangetal:2013}. One study has even demonstrated that the exposure to PM$_{2.5}$ can reduce the human life span by about 8.6 months \citep[][]{Krewski:2009}. \cite{Lelieveldetal:2015} concluded that the worldwide outdoor air pollution, mostly by PM$_{2.5}$, may result in over 3 million premature mortalities annually, mostly happening in Asia. 

The prediction of PM$_{2.5}$ concentrations can help minimize exposure time and reduce the risk of cardiovascular and respiratory deceases. There are commonly two types of models to predict the PM$_{2.5}$ concentrations: chemical-physical based models and statistical based models. The former models simulate the complex chemical reactions and physical dispersion processes and predict the PM concentration after a series of chemical reactions, transport and deposition, and the latter usually does not involve any chemical or physical mechanism and predict the PM concentrations after learning the past behaviors of air pollution and/or its relationships with some meteorological factors.   

The chemical-physical based models, also called chemical transport models (CTMs), such as CMAQ, GEOS-Chem, LOTOS-EUROS, MOZART CLaMS, \\ CAMx, MATCH, SKIRON, NAME, MOCAGE can interpret the mechanisms that control the PM$_{2.5}$ concentration \citep[][]{Brasseuretal:1998,Morris2002,McKennaetall:2002,Fuscoetal:2003,Dufouretal:2005,Langneretal:2005,Tescheetal:2006,Jonesetal:2007,Schaapetal:2008,Spyrouetal:2010}. However, due to its complexity,  these models usually require intensive computation and significant approximations \citep[][]{Cobourn:2010} and the accuracy of the prediction is controlled by the boundary layer schemes and the input data quality \citep[][]{Isukapalli:1999,Hanetal:2008}. The results of these models usually have high spatial and temporal resolutions.

However, because of simple implementation and fast processing, statistical models are widely used in predicting the air quality. Commonly used statistical models include artificial neural network (ANN) \citep[see e.g.,][]{ChanandLe:2013}, regression models \citep[see e.g.,][]{Cobourn:2010}, hidden Markov models (HMMs) \citep[see e.g.,][]{Sunetal:2013}. Compared to chemical-physical based models, results of statistical models can only reflect an average scenario of a certain place. 
  
Traditional models are usually used to predict the air pollution of a single urbanized area \citep[see e.g.,][]{Perezetal:2000}. However, the overall air pollution of a large area with many populated cities might be worth studying since sometimes we may need to evaluate the overall severity of the air pollution for a portfolio of cities, or to make comparisons with other groups of cities. 

Currently, most of previous studies on risk induced by air pollution are normally only focused on health risk to humans \citep[see e.g.,][]{PopeIIIetal:2002}, however, since clean air sometimes can be treated as a natural resource and therefore has economic values \citep[][]{FreemanIIIetal:2014}. The air may have the risk of devaluation due to air pollution. However, to our knowledge, there is very limited research on risk assessment on the value of clean air.

Therefore, one of the purposes of this paper is to explore a method to examine the overall severity of the air pollution for a portfolio of cities and propose a method to evaluate the clean-air-at-risk. There are several studies which analyze air pollution data using copula model. \cite{garcia2013exploring} use copula functions to model the dependence between wind direction and SO$_2$ concentration. \cite{noh2013copula} estimate a regression function based on copulas. \cite{zhanqiong2013modeling} use copula based GARCH models to capture the dependence structure between Air Pollution Index of Shenzhen and regional levels. To our knowledge, although the dependence structure of pollutants at the same location or pollution levels between different locations has been estimated using copula functions, it has never been used for understanding the overall severity of the air pollution for a portfolio of cities. 

Another purpose of this paper is to propose a state-of-the-art method to predict the PM$_{2.5}$ levels based on historical PM$_{2.5}$ and weather data. As we mentioned, ANN is one of the commonest statistical method to predict air pollution. We will implement a long short-term memory (LSTM) machine learning structure that has not been used in air quality forecast before and compare its performance with other traditional methods, such as logistic regression and multilayer perceptrons.

In this paper we first fit a copula dependence structure between PM$_{2.5}$ concentrations of five cities (Beijing, Tianjin, Hengshui, Chengde, and Xingtai) in Beijing-Tianjin-Hebei area based on daily PM$_{2.5}$ data. Then we show how to efficiently simulate the risk measures clean-air-at-risk and conditional clean-air-at-risk under the $t$-copula model. (Efficient simulations are important as the number of cities might be too big.) Second, we will compare the performance of three statistical based models (logistic regression, a two hidden-layer perceptron, and LSTM) in predicting the PM$_{2.5}$ levels at 3 hour scale for the four cities (Beijing, Chengde, Xingtai, and Zhangjiakou) based on the 3-hourly historical PM$_{2.5}$ and meteorological data. This part of the paper diverges from the literature on the prediction of PM$_{2.5}$ levels by implementing long short-term memory (LSTM) machine learning architecture. In contrast to multilayer perceptrons, LSTM tries to link current pollution levels to not only the features of current time but also previous time points.

\section{Methodology}
\subsection{The overall PM$_{2.5}$ pollution risk simulation}

\subsubsection{Data collection and preparation}
\begin{figure}[h]
 \includegraphics[width=.5\textwidth]{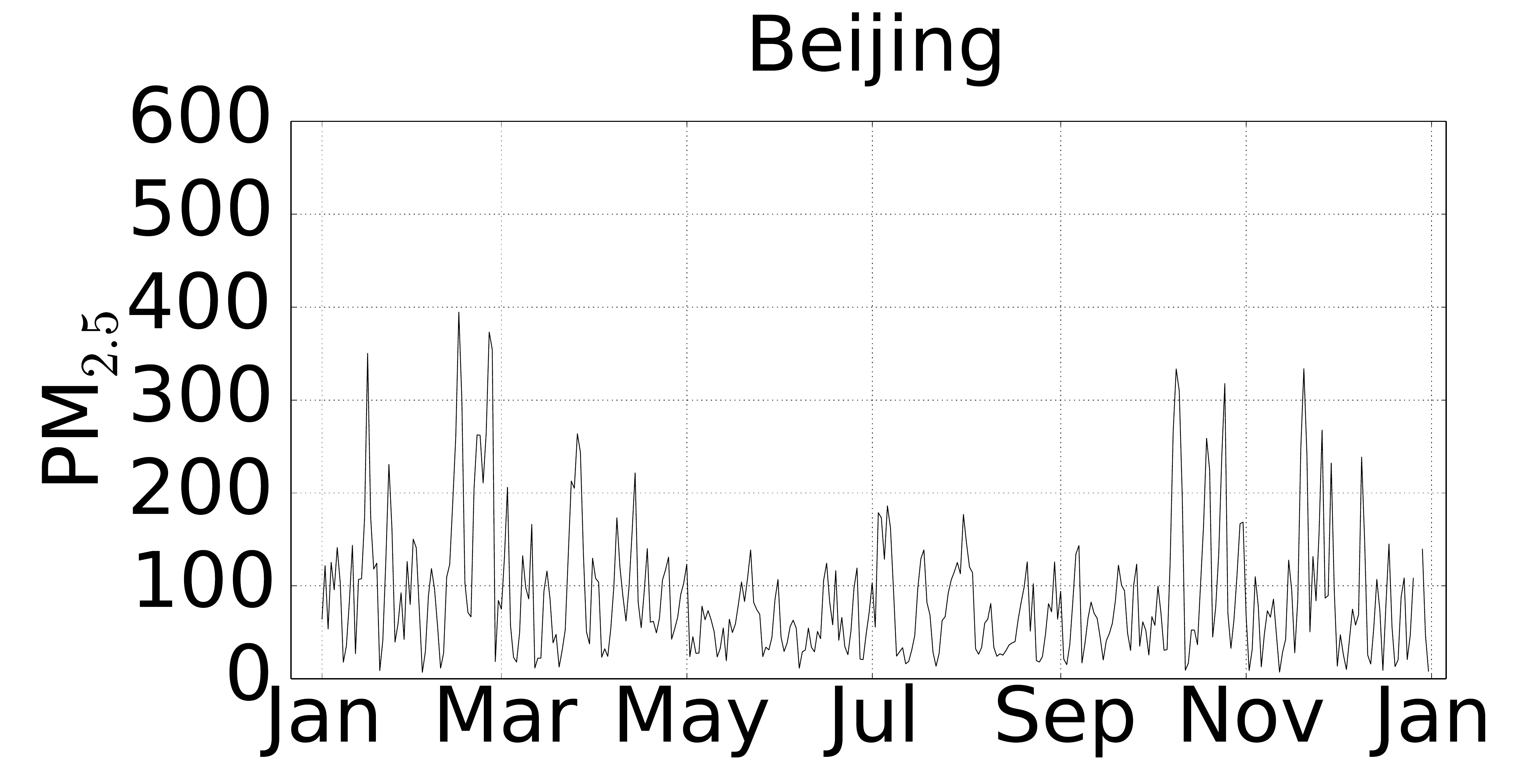}~
 \includegraphics[width=.5\textwidth]{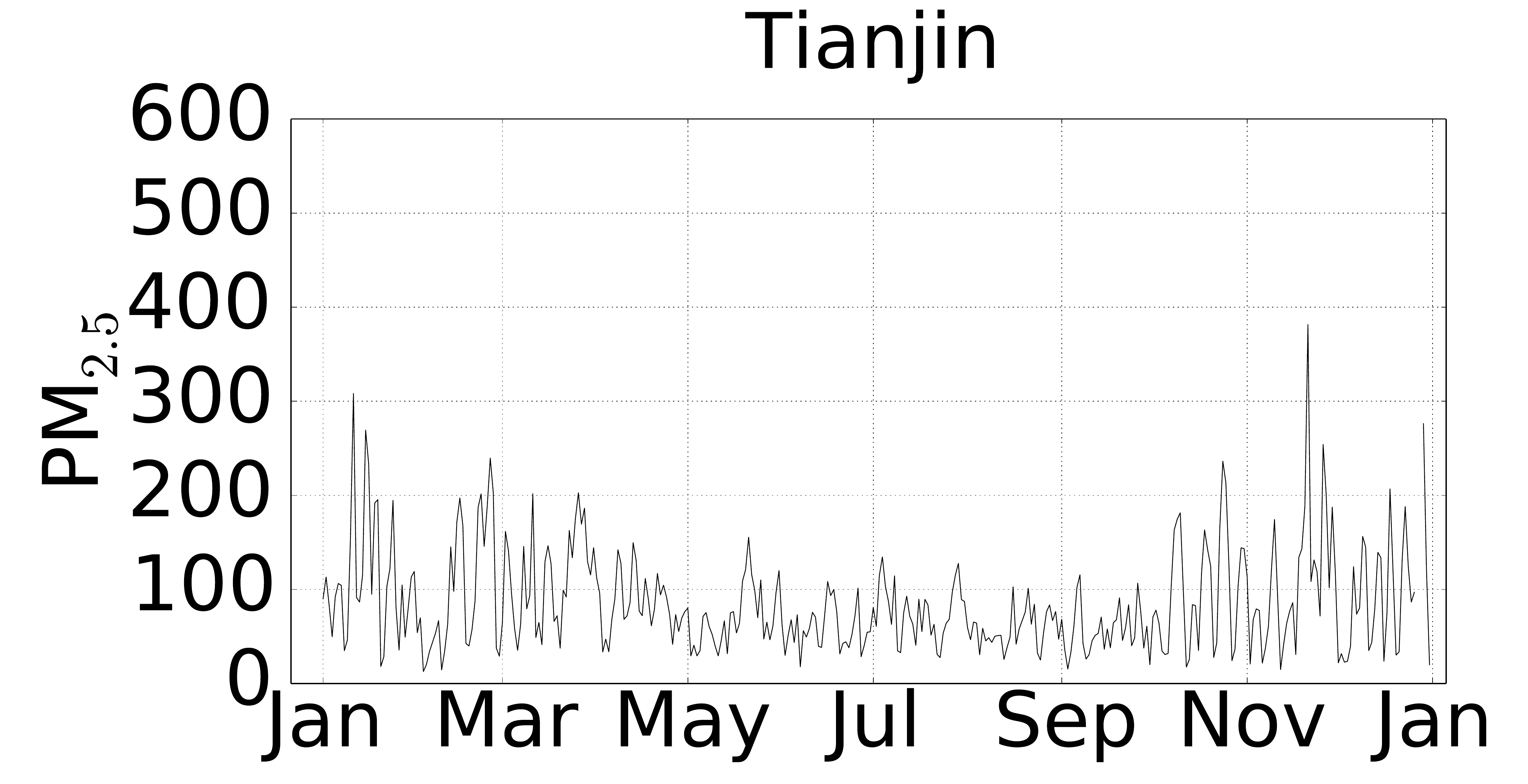}\\
 
 \includegraphics[width=.5\textwidth]{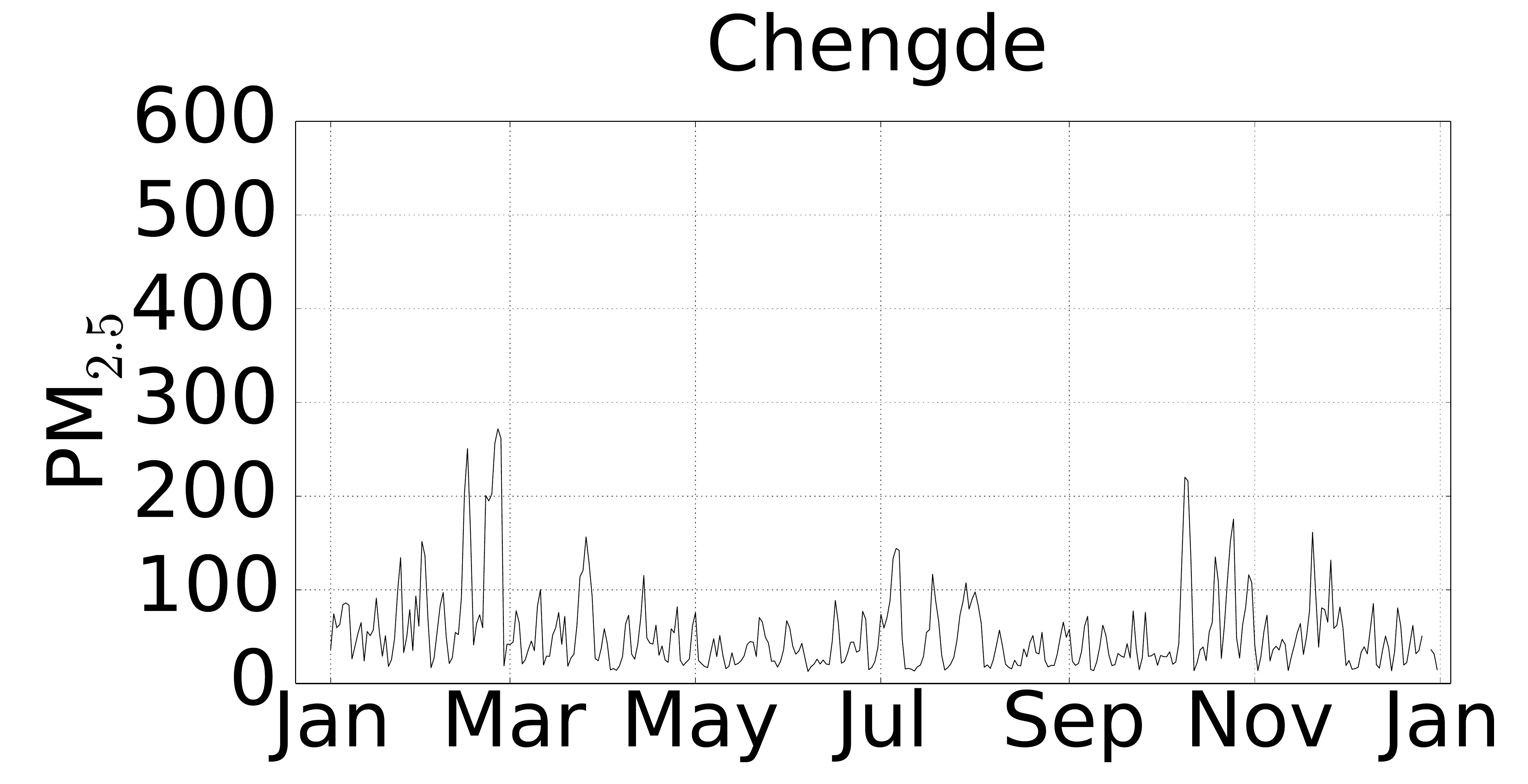}~
 \includegraphics[width=.5\textwidth]{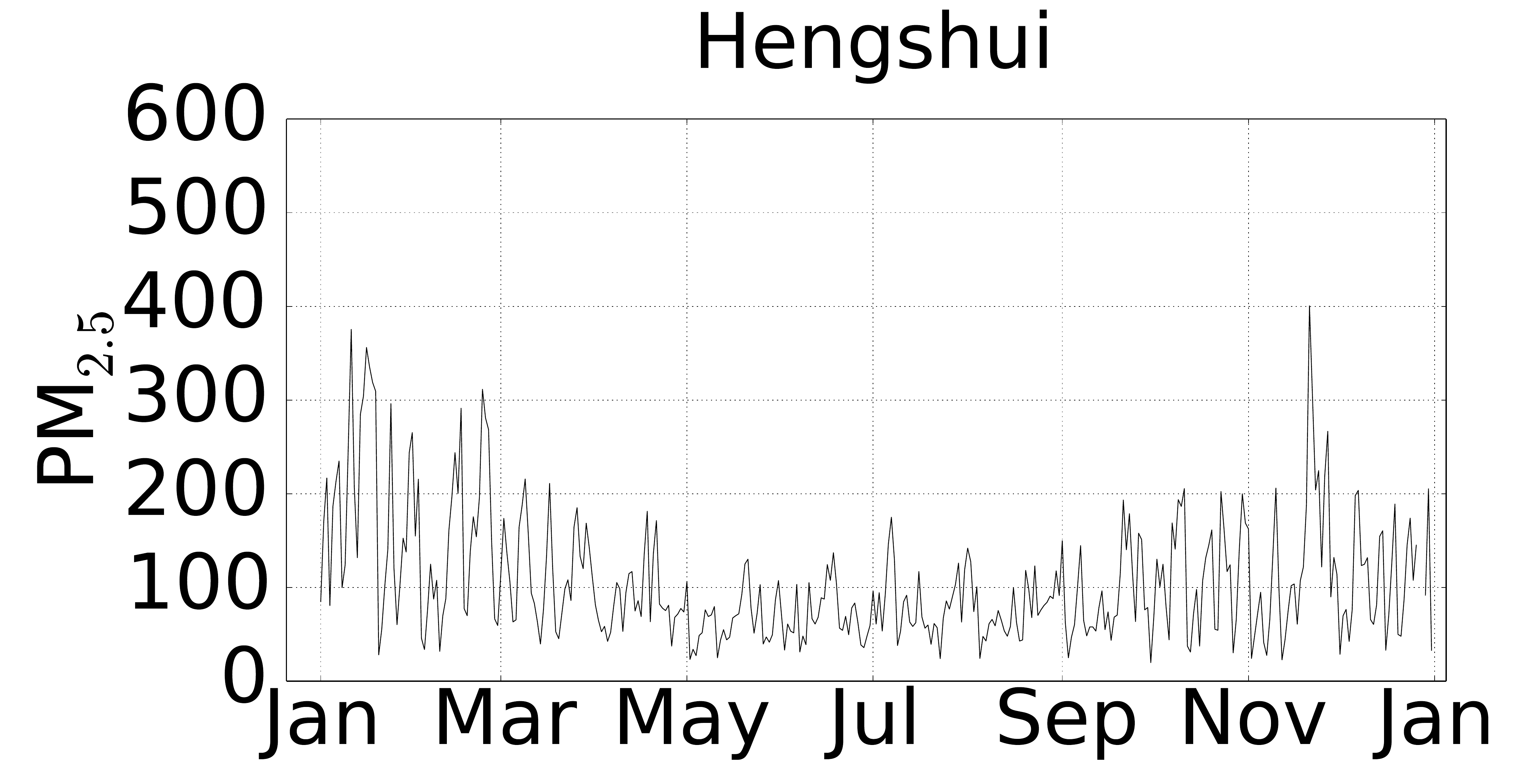}\\
 
 \includegraphics[width=.5\textwidth]{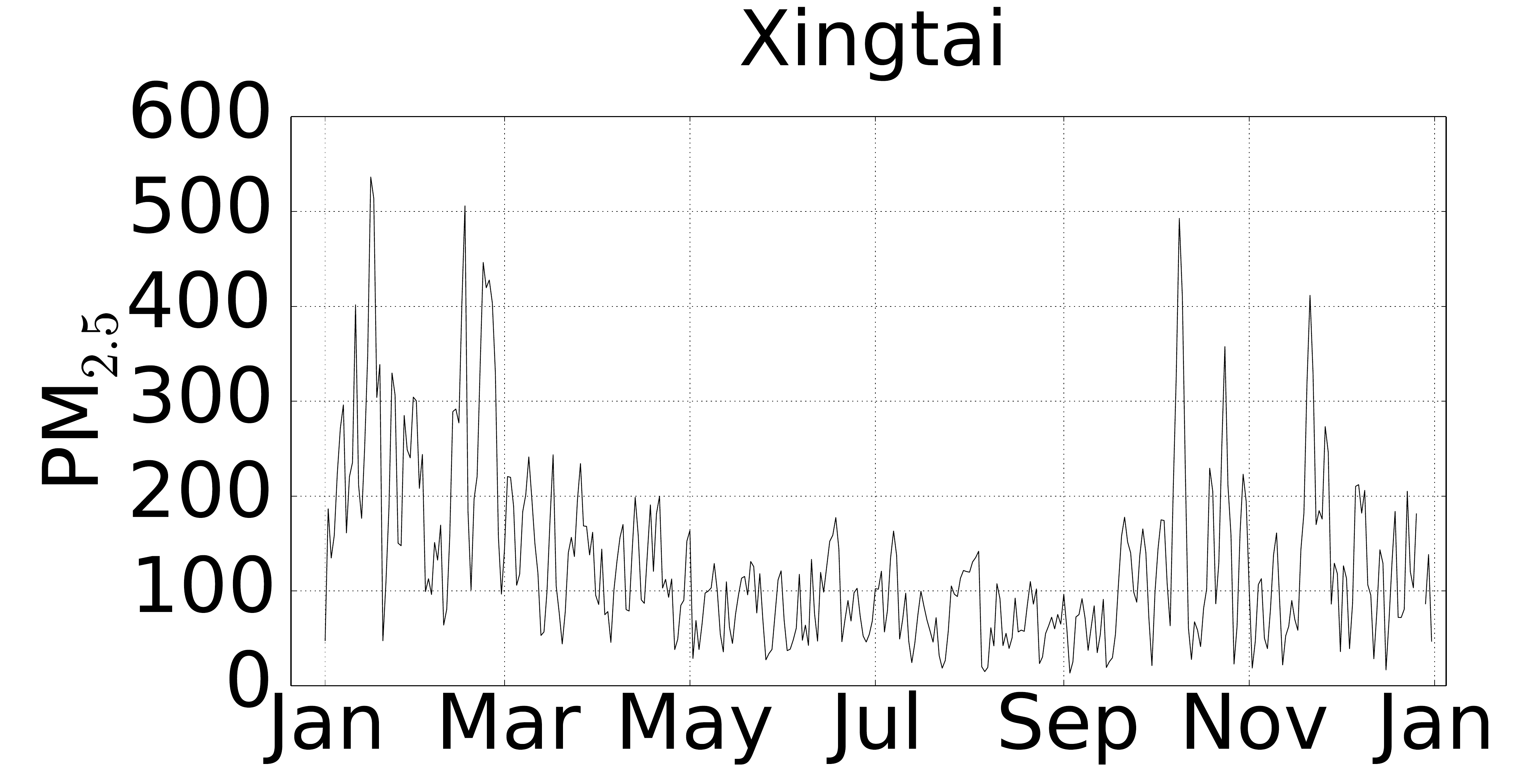}
 \caption{The daily PM$_{2.5}$ concentrations in 2014. \label{fig:PM2.5}}
 \label{pm25Plots}
\end{figure}

Since Beijing-Tianjin-Hebei is one of the most polluted area in China, we selected five cities (Beijing, Tianjin, Chengde, Hengshui, and Xingtai) to evaluate the overall severity of PM$_{2.5}$ pollution and clean-air-at-risk. Daily PM$_{2.5}$ concentrations published by China National Environmental Monitoring Center were collected for the whole year of 2014, with exceptions of days 361 and 362 on which the data was not collected. Daily PM$_{2.5}$ concentrations in Beijing (Bj), Tianjin (Tj), Chengde (Cd), Hengshui (Hs), and Xingtai (Xt) for the year 2014 are shown in Figure~\ref{pm25Plots}. Due to the limitation of data source, we only selected these five cities to demonstrate the methodology.

\subsubsection{Model}
The overall PM$_{2.5}$ concentration of a portfolio of cities at time $0$ can be written as: 
\begin{equation}
\text{C}_0=\sum_{d=1}^{D}w_{d}\,\text{PM}^{0}_{d},
\end{equation}
where $w_{d}$ is the population of city $d$, PM$^{0}_{d}$ is the PM$_{2.5}$ concentration of city $d$ at time $0$, and $D$ is the number of cities, which equals to 5 in our numerical experiments. We selected population as the weighting term because the severity of air pollution is associated with population. Cities with high PM levels but small population may contribute only a little to the overall PM$_{2.5}$ pollution. 

\cite{sak;2011} first fit a copula dependence structure for the log returns of commodity metals then implement an efficient simulation method for simulating the supply portfolio risk. The same portfolio approach can be applied to the overall pollution risk. We first define log of PM$_{2.5}$ ratio over a one day horizon for $d$th city, $r_{d}$, as
 \begin{equation}
 r_{d}=\log\left(\frac{\text{PM}^{1}_{d}}{\text{PM}^0_{d}}\right),
 \label{log-ratio}
\end{equation}
where PM$^{1}_{d}$ is the PM$_{2.5}$ concentration of city $d$ for the next day.

We assume that the log-ratio of $D$ cities, $r_{d},\,d=1,\cdots,D,$ over a day follow the normal or $t$-copula and its dependence structure is described by the positive definite matrix, $\Sigma$. $L$ denotes the (lower-triangular) Cholesky factor of $L$ satisfying $LL'=\Sigma$. The classical random return vector generation algorithm from the normal and t-copula starts with a vector $Z$ of $D$ independent and identically distributed standard normal variates and then transforming it into the correlated normal vector $\widetilde{Z}=LZ$. For the normal copula, $V=\widetilde{Z}$, and for the $t$-copula, a random variate $Y$ from a chi-squared distribution with $\nu$ degrees of freedom ($\chi_{\nu}^{2}$) is generated to obtain the random vector $V=\widetilde{Z}/\sqrt{Y/\nu}$. Then, the log-ratio vector $r=(r_{1}, r_{2},\cdots, r_{D})'$, can be written as a function of the random input vector $V$:
\begin{equation}
r_{d}(V_{d})=s_{d}\,G_{d}^{-1}(F(V_{d})),
\end{equation}
where $F$ denotes the cumulative distribution function (CDF) of a standard normal distribution for the normal copula and the CDF of a $t$-distribution with $\nu$ degrees of freedom for the $t$-copula. $G_{d}$ denotes the CDF of the marginal distribution of the PM$_{2.5}$ log-ratio of $d$th city and $s_{d}$ denotes the scaling factor related to the daily volatility, $\sigma_{d}$, and the variance, $var_{d}$, of the $d$th marginal distribution is given by
\begin{equation*}
s_{d}=\sigma_{d}\sqrt{\frac{1}{var_{d}}}.
\end{equation*}
Then, given that PM$_{2.5}$ concentrations of cities at time 0 are available, the overall PM$_{2.5}$ concentration for the next day is:
\begin{equation}
C = \sum_{d=1}^{D}w_{d}\,\text{PM}_{d}^{0}\,e^{s_{d}G_{d}^{-1}(F(V_{d}))}.
\end{equation}

Value-at-risk (VaR) and conditional value-at-risk (CVaR) are widely used risk measures on financial portfolios. We define two similar measures; clean-air-at-risk (CaR) and conditional clean-air-at-risk (CCaR). For a given city portfolio and a probability level $\alpha$, the CaR$_{\alpha}$ is defined as the smallest number $\tau$ such that the probability of $C$ exceeds $\tau$ is at most $1-\alpha$, and CCaR$_{\alpha}$ denotes the conditional expectation of $C$ given $C > \text{CaR}_{\alpha}$. The CaR$_{\alpha}$ and CCaR$_{\alpha}$ for a probability level $\alpha$ can be represented as follows:
\begin{eqnarray*}
\text{CaR}_{\alpha}&=&\inf\{\tau:P(C>\tau)\leqslant 1-\alpha\},\\
\text{CCaR}_{\alpha}&=&E[C|C>\text{CaR}_{\alpha}].
\end{eqnarray*}

\subsubsection{Model parametrization}

We use the inference functions for margins method \citep[see e.g.,][]{malevergne;2006} to fit a dependence structure between PM$_{2.5}$ daily log-ratios of cities (given in (\ref{log-ratio})). We estimate  the parameters of marginal distributions and the copula using likelihood maximization sequentially in the given order. Ninety percent of the data were chosen randomly to fit the model. The rest of the data are used for measuring the goodness of the fit of the marginal distributions.

The correlation matrix of the log-ratios are given in Table~\ref{correlTable}. The maximum linear correlation is 0.814, which is between Beijing and Chengde. The minimum linear correlation is 0.459, which is between Chengde and Hengshui. This table clearly shows that there is a strong dependence between log-ratios of cities.

\begin{table}[h]
\caption{Correlation matrix of the PM$_{2.5}$ log-ratios.}
\centering
\begin{tabular}{cccccc}
\hline
City&Bj&Tj&Cd&Hs&Xt\\ \hline
Bj & 1.000 & 0.753 & 0.814 & 0.553 & 0.599\\
Tj & 0.753 & 1.000 & 0.627 & 0.749 & 0.665\\
Cd & 0.814 & 0.627 & 1.000 & 0.459 & 0.502\\
Hs & 0.553 & 0.749 & 0.459 & 1.000 & 0.766\\
Xt & 0.599 & 0.665 & 0.502 & 0.766 & 1.000\\ \hline
\end{tabular}
\label{correlTable}
\end{table}

Our extensive numerical analysis show that the best fitting model for the PM$_{2.5}$ log-ratios of the cities is the t-copula dependence structure with a degrees of freedom of $11.78\, (\nu)$ and the generalized hyperbolic marginals. The fitted marginal distributions and correlation matrix of the $t$-copula (with standard errors of the point estimates) are given in Tables~\ref{margfits} and \ref{tcopulaSigma}.

\begin{table}[h]
\caption{Parameters of the fitted generalized hyperbolic marginals for the PM$_{2.5}$ log-ratios.}
\centering
\begin{tabular}{cccccc}
\hline
City & $\lambda$ & $\alpha$ & $\delta$ & $\beta$ & $\mu$ \\ \hline
Bj & 0.1894 & 2.4296 & 0.7561 & -1.0516 & 0.5075 \\
Tj & 1.8041 & 3.3702 & 0.0066 & -0.8673 & 0.2959 \\
Cd & 1.1848 & 6.4420 & 0.5492 & -4.0233 & 0.7318 \\
Hs & 1.7675 & 4.8022 & 0.4498 & -1.7954 & 0.4339 \\
Xt & 2.0100 & 3.9889 & 0.0500 & -1.0875 & 0.3041 \\ \hline
\end{tabular}
\label{margfits}
\end{table}

\begin{table}[h]
\caption{Correlation matrix of the fitted $t$-copula for the PM$_{2.5}$ log-ratios. Standard errors of the point estimates are given in parentheses.}
\centering
\begin{tabular}{cccccc}
\hline
City&Bj&Tj&Cd&Hs&Xt\\\hline
Bj & 1.000 & 710 & 0.744 & 0.487 & 0.577\\
   &           &(0.024)&(0.022)&(0.039)&(0.034)\\
Tj & 0.710 & 1.000 & 0.549 & 0.709 & 0.623\\
   &           &           &(0.035)&(0.024)&(0.0331)\\
Cd & 0.744 & 0.549 & 1.000 & 0.382 & 0.463\\
   &           &           &           &(0.044)&(0.041)\\
Hs & 0.487 & 0.709 & 0.382 & 1.000 & 0.729\\
   &           &           &           &          &(0.023)\\
Xt & 0.577 & 0.623 & 0.463 & 0.729 & 1.000\\\hline
\end{tabular}
\label{tcopulaSigma}
\end{table}

\subsubsection{Efficient simulation of pollution risk}

This section explains how to efficiently simulate pollution risk under the $t$-copula model. We employ the importance sampling and stratified importance sampling methods, which will be briefly summarized. But, before that, we start with the naive Monte Carlo simulation method. 

Here we give the methodology for simulating the exceeding probability (EP) 
\[
P(C>\tau) = E[\textbf{1}_{\{C>\tau\}}],
\]
where $\textbf{1}_{\{.\}}$ denotes the indicator function and conditional excess (CE)
\begin{equation*}
E[C|C>\tau]=\frac{E[C\textbf{1}_{\{C>\tau\}}]}{E[\textbf{1}_{\{C>\tau\}}]}.
\end{equation*}
The risk measure CaR$_{\alpha}$ can be calculated by inverting probability distribution of exceeding probability for a probability level of $\alpha$. Then, CCaR$_{\alpha}$ can be simulated using $\tau=\text{CaR}_{\alpha}$.

Each replication ($k=1,\cdots,N$) of the naive Monte Carlo algorithm for simulating EP and CE follows the steps given below:
\begin{enumerate}
	\item[1.] Generate D standard normal random variables, $Z = (Z_{1}, \cdots, Z_{D})'$, and a chi-squared random variable $Y$ with $\nu$ degrees of freedom, all independently.
	\item[2.] Calculate the random vector $V=\widetilde{Z}/\sqrt{Y/\nu}$.
	\item[3.] Calculate $C^{(k)}=\sum_{d=1}^{D}w_{d}\text{PM}^{0}_{d}\,e^{s_{d}G_{d}^{-1}(F(V_{d}))}$ for $k$th replication.
\end{enumerate}
After computing $N$ replications of $C$, the naive Monte Carlo simulation estimates of EP and CE can be calculated using:
\begin{equation*}
\text{EP}^{\text{NV}} = \sum_{k=1}^{N}\textbf{1}_{\{C^{(k)}>\tau\}},
\end{equation*}
and
\begin{equation*}
\text{CE}^{\text{NV}} = \frac{\sum_{k=1}^{N}C^{(k)} \textbf{1}_{\{C^{(k)}>\tau\}}}{\sum_{k=1}^{N}\textbf{1}_{\{C^{(k)}>\tau\}}}.
\end{equation*}

When the threshold value $\tau$ is large (in the rare event setting), most of the naive simulation replications return zero. This increases the variance of the simulation. To remedy this problem, importance sampling (IS) can be used to modify the joint density of the random input. We use the IS technique described in \cite{Sak:Hormann:Leydold;2010a}. We add a mean shift vector $\mu$ with positive entries to the normal vector $Z$ and use a scale parameter $\theta$ less than two for the chi-square (i.e. gamma) random variable $Y$.

The IS estimates for EP and CE are as follows: 
\begin{eqnarray*}
P(C>\tau)&=&\tilde{E}[\textbf{1}_{\{C>\tau\}}W_{\mu,\theta}(Z,Y)],\\
E[C|C>\tau]&=&\frac{\tilde{E}[C\textbf{1}_{\{C>\tau\}}W_{\mu,\theta}(Z,Y)]}{\tilde{E}[\textbf{1}_{\{C>\tau\}}W_{\mu,\theta}(Z,Y)]},
\end{eqnarray*}
where $W_{\mu,\theta}(Z,Y)$ is the likelihood ratio under the IS density and it is calculated as
\begin{equation*}
W_{\mu,\theta}(Z,Y)=\exp(-\mu'Z+\mu'\mu/2-Y/2+Y/\theta+\log(\theta)\nu/2).
\end{equation*}
For more details on the determination of the IS parameters and implementation of the simulation algorithm, see Section 4 of \cite{Sak:Hormann:Leydold;2010a}.

To obtain further variance reduction, one can stratify the importance sampling density along one or possibly more directions \citep[see][]{bacsouglu;2013}. For the $t$-copula model, suppose that $\xi_i$, $i = 1, \ldots , I$, is a partition of $\mathcal{R}^{D+1}$ into $I$ disjoint subsets with probabilities $\tilde p_i=\tilde P\left(\left(Z,Y\right) \in \xi_i\right)$ under the IS density. Then the stratified importance sampling (SIS) estimates for EP and CE are as follows: 
\begin{eqnarray*}
P(C>\tau)&=&\sum_{i=1}^{I}\tilde{p}_{i}\tilde{E}[\textbf{1}_{\{C>\tau\}}W_{\mu,\theta}(Z,Y)|(Z,Y)\in\xi_{i}],\\
E[C|C>\tau]&=&\sum_{i=1}^{I}\tilde{p}_{i}\frac{\tilde{E}[C\textbf{1}_{\{C>\tau\}}W_{\mu,\theta}(Z,Y)|(Z,Y)\in\xi_{i}]}{\tilde{E}[\textbf{1}_{\{C>\tau\}}W_{\mu,\theta}(Z,Y)|(Z,Y)\in\xi_{i}]}.
\end{eqnarray*}

To minimize the variance of the stratified estimator, adaptive optimal allocation (AOA) algorithm is applied in each iteration. AOA modifies the proportion of the replications of each strata based on the estimation of conditional standard deviation. For more details on AOA and SIS, see Sections 2 and 5 of \cite{bacsouglu;2013}.

\subsubsection{Simulation results}

In order to illustrate the efficiency of the IS and SIS methods, we implement all algorithms in R \citep{rcoreteam}. For the model parameters, we use the fitted parameters given in Tables~\ref{margfits} and \ref{tcopulaSigma}. We use $w=\{0.4132, 0.2726, 0.0732, 0.0914,$ $0.1496\},$ for Bj, Tj, Cd, Hs, and Xt in the given order, $s_d=1$, for $d=1,\cdots,D$ and initial PM$_{2.5}$ concentrations of cities are assumed to be $100 \,\SI{}{\micro\gram\meter^{-3}}$ ($\text{PM}_{d}^{0}=100$).

We present CaR$\alpha$, CCaR$_{\alpha}$ values and 95\% confidence intervals as percentage of the point estimates for the naive, IS and SIS methods over a one-day horizon in Table~\ref{simulRes}. In these experiments, the total number of replications used is $N=10^5$ for naive and IS simulations and approximately $N\approx 10^5$ for SIS simulation. We terminate SIS in four iterations, using approximately 10, 20, 30, and 40 percent of the total sample size in each iteration, sequentially. Variance reduction (VR) factors (greater than 1 means that there is variance reduction compared to the naive method) indicate the relative efficiency of the IS and SIS with respect to the naive simulation in computing CCaR$_{\alpha}$. 
 
\begin{table}[h]
\caption{CaR$\alpha$, CCaR$_{\alpha}$ values and 95\% confidence intervals as percentage of the point estimates for the naive, IS and SIS methods over a one-day horizon. Variance reduction factors for the IS and SIS are also provided.}
\centering
\resizebox{12cm}{!}{
\begin{tabular}{cccccccccc}
\hline
 & & &Naive&& \multicolumn{2}{c}{IS}&& \multicolumn{2}{c}{SIS}\\ \cline{4-4} \cline{6-7} \cline{9-10}
 && &Confidence&&Confidence&VR&&Confidence&VR\\
$\alpha$&CaR$_{\alpha}$&CCaR$_{\alpha}$&interval&&interval&factor&&interval&factor\\
\hline
0.05 & 239.32 & 315.34 & $\pm0.91\%$ &&  $\pm0.23\%$ &16&&$\pm0.06\%$&201\\
0.01 & 352.03 & 461.16 & $\pm2.01\%$  && $\pm0.24\%$ &67&&$\pm0.08\%$&650\\
0.005 & 414.22 & 543.20 & $\pm2.93\%$ && $\pm0.25\%$ &134&&$\pm0.08\%$&1339\\
0.002 & 515.27 & 677.76 & $\pm4.80\%$ && $\pm0.26\%$ &337&&$\pm0.09\%$&3055\\
0.001 & 600.78 & 791.60 &  $\pm6.28\%$ && $\pm0.27\%$ &524&&$\pm0.09\%$&5246\\
\hline
\end{tabular}}
\label{simulRes}
\end{table}

\begin{figure}[htbp]
 \centering
 \includegraphics[width=.8\textwidth]{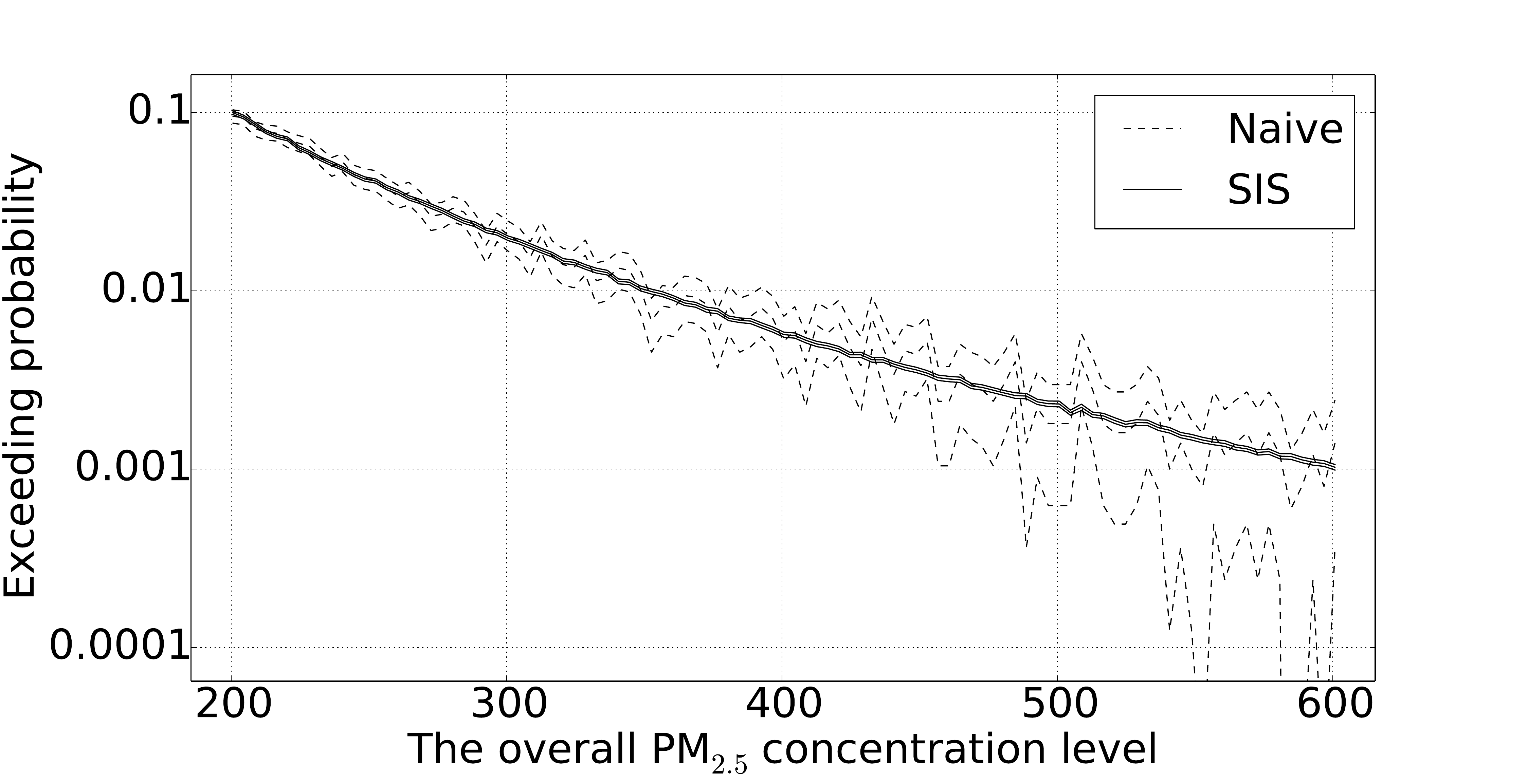}
 \caption{Exceeding probabilities of the overall PM$_{2.5}$ concentration for time horizon of one day using the SIS method (with approximately 5000 replications) and the naive method (with 5000 replications). \label{fig:simu}}
\end{figure}

To give a rough idea of how EP changes with respect to the overall PM$_{2.5}$ concentration threshold, we simulate EP for various threshold values in Figure~\ref{fig:simu}. Point estimates and also 95\% confidence intervals for the naive and SIS methods are provided. The efficiency of the SIS method can be easily observed by comparing the widths of the confidence intervals. The naive method gives wider confidence intervals and stops giving sensible confidence intervals for thresholds greater than 540.

\subsection{Prediction of PM$_{2.5}$ levels}
\subsubsection{Data Collection}

We collect pollution (PM$_{2.5}$) and weather data of four Chinese cities; Beijing, Chengde, Xingtai, and Zhangjiakou at three-hour time intervals for 2014. The data source is China National Environmental Monitoring Center and NOAA. Our time series for each city have 2919 points. A  descriptive statistics for the features that we use in our machine learning models are given in Table~\ref{tabDS}. Some of the features have missing values at some time points. The number of missing values are given in ``\#NA" columns for all the cities in Table~\ref{tabDS}. Note that s.e. stands for standard error on mean in Table~\ref{tabDS}.

\begin{table}[htbp]
\caption{Statistics of features for Beijing, Chengde, Xingtai, and Zhangjiakou for 2014.}
	\centering 
	\resizebox{12.5cm}{!}{
\begin{tabular}{lllccccccccccc}
\hline
              & & &\multicolumn{2}{c}{Beijing} & &\multicolumn{2}{c}{Chengde} && \multicolumn{2}{c}{Xingtai} && \multicolumn{2}{c}{Zhangjiakou} \\ \cline{4-5} \cline{7-8} \cline{10-11} \cline{13-14}
  Feature  & Unit & Range& Mean (s.e.)&\#NA&& Mean (s.e.)&\#NA&& Mean (s.e.)&\#NA&& Mean (s.e.)&\#NA \\ \hline
PM$_{2.5}$ & \SI{}{\micro\gram\meter^{-3}}  & [3.4,674.0] &88.7 (1.4)& 29& & 53.6 (0.9)& 147 & & 129 (2.0) & 145 & & 34.7 (0.9) & 150\\
Temperature & \SI{}{\celsius}  &[-19.7,40.7]  & 14.2 (0.2)&8 &  & 9.9 (0.2)& 22& & 15.6 (0.2) &19 & & 9.6 (0.2) & 17\\
Humidity & \SI{}{\percent}  & [7,100]  &51.0 (0.4)&8 & &53.4 (0.5) & 24 & & 57.4 (0.4) & 19 & & 45.7 (0.4) & 17\\
Pressure & \SI{}{\milli \bar}  &[994-1044] &1017 (0.2)&8 & & 1018 (0.2) &23  & &1017 (0.18)& 23 & &1018 (0.2) & 19\\
Wind speed & \SI{}{\meter\per\second}  & [0,32.4]  &2 (0.03)& 109 & &2 (0.03)  & 372  & & 1.6  (0.02) & 243 & & 3.0 (0.04) & 160\\
Wind direction & \SI{}{\degree}  &[0,360] &163.7 (2.0)& 109& & 202 (2.3) & 372 & & 164.9 (2.0) & 243 & & 225.7 (2.0) & 160 \\
Hour of Day & int  & [2,23] & - & - &  & -  & - & & - & - & &- &- \\
\hline
\end{tabular}
}
\label{tabDS}
\end{table}

We combine the features and the level that we want to forecast at each time point to construct the data. The number of rows which contains a missing value or values in any columns (features and level) are 144, 587, 468, and 395 for Beijing, Chengde, Xingtai, and Zhangjiakou in the given order out of 2918 (last row of the data does not have level thus we remove it) rows for each city.

\subsubsection{Model}
\begin{figure}[h]
\begin{center}
\includegraphics[width=.8\textwidth]{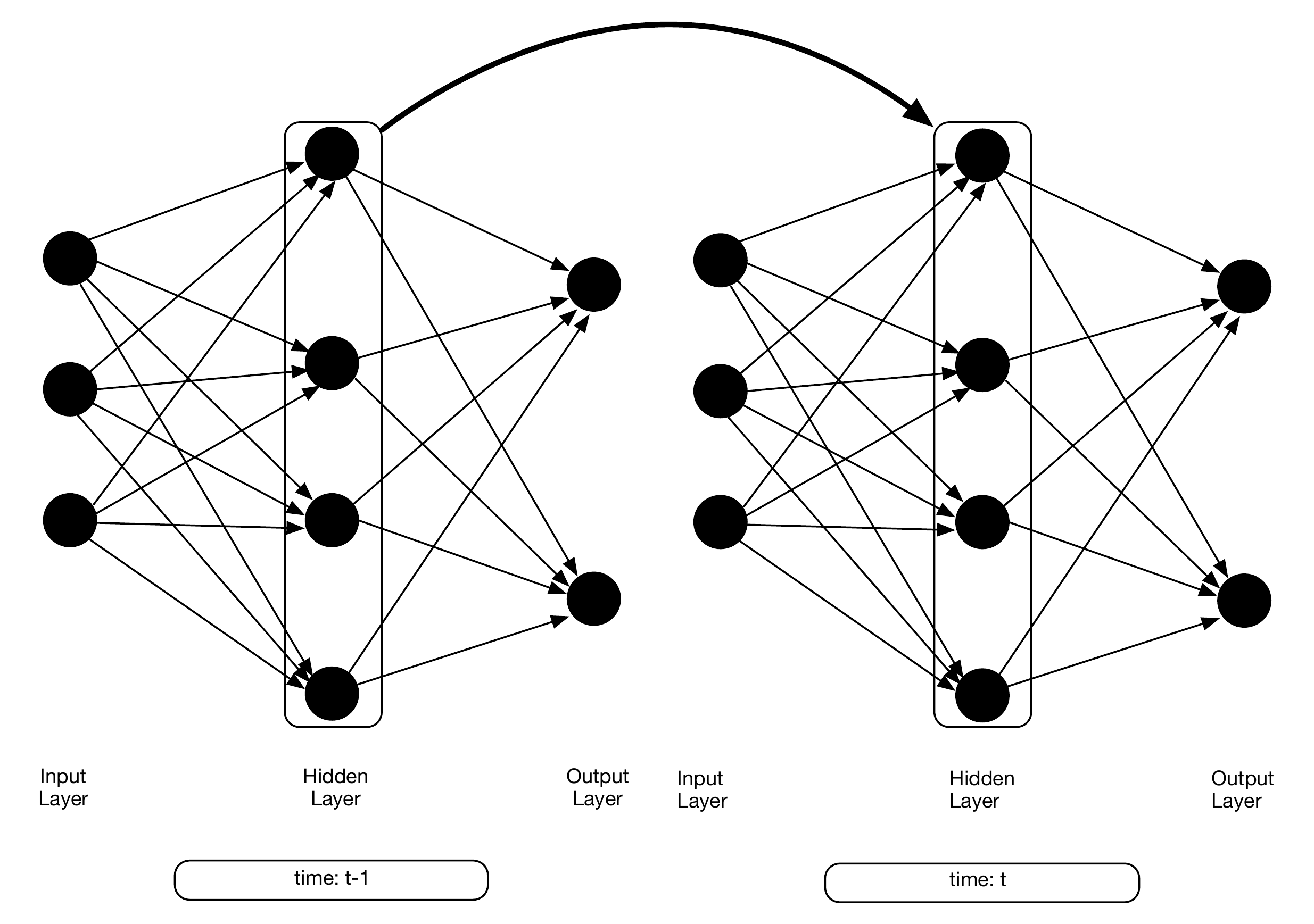}
\caption{Artificial Neural Network}
\label{neuralNetw}
\end{center}
\end{figure}

This section briefly introduces machine learning techniques implemented in this study. An Artificial neural network (ANN) is simply a connection of nodes by weights (see Figure~\ref{neuralNetw}). If no cycles are allowed, this kind of ANNs are known as feedforward neural networks. Recurrent neural networks, on the other hand, have cyclic connections. 

We first  describe multilayer perceptrons, most widely used form of feedforward neural network. Then, long short-term memory recurrent neural networks are discussed. For more information on machine learning techniques see e.g., \cite{Graves;2008}, \cite{Sutskever;2013}, and \cite{Bengio-et-al-2015-Book}.

A multilayer perceptron (no recurrent arcs in Figure~\ref{neuralNetw}) starts with an input layer and activations (generally a nonlinear function is applied to a weighted sum of prior nodes' activations) are propagated to the output layer with feed forward connections. We might have multiple hidden layers in between the input and output layer. As we don't have any connecting arcs between nodes at different time points, allocation of the data to training and test sets can be randomized.

The neural networks at different time points share the same weights. Network weights are found by minimizing a loss function (increasing the likelihood of observations) on training data set. For the optimization, stochastic gradient or a more complex method like L-BFGS \citep[][]{Byrd;Lu;Nocedal;Zhu:1995a} can be employed. As all these methods require the first derivatives of loss function with respect to network weights, a backward pass (computing derivatives using chain rule) of the network is necessary. Thus, training a multilayer perceptron is simply a repetition of forward and backward passes of the network while updating the network weights using the first derivatives. 

In the recurrent neural networks, hidden layer activations are also affected from the hidden layers of the previous time points as shown in Figure~\ref{neuralNetw}. Thus, each output in the network might be also a function of previous inputs. Training of recurrent neural networks is not easy due to well documented problems, exponentially decaying or exploding derivatives \citep[see][]{Hochreiter;1991,Bengioetal:1994}. Long short-term memory (LSTM) architecture provides a solution for decaying derivatives problem \citep[see][]{Hochreiter;Schmidhuber:1997}. LSTM architecture includes memory units to be able to remember and retrieve important information over long periods of time. LSTMs have been successfully applied to many fields like speech recognition, robotic control etc. \citep[see e.g.,][]{Hochreiter;Schmidhuber:2005, Mayeretal:2006}

LSTM along with other machine learning algorithms have been implemented in a scientific computing framework named \textit{Torch}. As it is reliable and easy-to-use, we used Torch7 \citep[see][]{Collobertetal:2011} to predict PM$_{2.5}$ levels. We employed three different methods, logistic regression (no hidden layer is used in the multilayer perceptron architecture), a two hidden-layer perceptron, and LSTM. 

\subsubsection{Numerical Results}

This section illustrates the performance of logistic regression (LR), a two hidden-layer perceptron (NN), and LSTM at predicting PM$_{2.5}$ levels (there are six levels of health concern) on the pollution data. We start with the task of trying to predict PM$_{2.5}$ levels of a single city using only its own features. We chose Beijing for this task since it has less missing values. For the implementation of LR and NN, we simply ignore data rows with missing values and allocate the rest of the data to training and test sets randomly. On the other hand, LSTM's implementation does the training on sequences which are not interrupted by missing values. (To our knowledge \textit{Torch7} has no functionality to work with missing values thus we programmatically solved this problem.) The classification accuracies of the machine learning methods for Beijing are given in the first two columns of Table~\ref{tabPP}. Note that, as the data size is small, the prediction powers are affected too much from the random selection of train and test sets. We report the average of 10-fold cross-validation classification accuracies for LR and NN in Table~\ref{tabPP}. For LSTM, we repeat 10 independent experiments and report the average of the classification accuracies. As the data size and number of features are small and only a small percentage of the data rows has missing values, all the methods performed similarly.

\begin{table}[htbp]
	\centering
\caption{Train and test classification accuracies of LR, NN, and LSTM on pollution data of Beijing, Chengde, Xingtai, and Zhangjiakou.}
\resizebox{12.5cm}{!}{
\begin{tabular}{ccccccccccccccc}
\hline
              & \multicolumn{2}{c}{Beijing} & &\multicolumn{11}{c}{Beijing, Chengde, Xingtai, and Zhangjiakou} \\ \cline{2-3} \cline{5-15}
              & \multicolumn{2}{c}{All features} & & \multicolumn{2}{c}{All features} && \multicolumn{2}{c}{Best} & &\multicolumn{2}{c}{NA Incl.} && \multicolumn{2}{c}{MSE Crit.}\\              
\cline{2-3} \cline{5-6} \cline{8-9} \cline{11-12} \cline{14-15}
& Train & Test& & Train & Test&& Train & Test& &Train & Test&& Train & Test \\ \hline
LR &   0.75& 0.74 & &  0.67 &0.66& &0.67&0.66&&0.65 & 0.64& &0.59 &0.59\\
NN&  0.81 & 0.76& & 0.78&0.75&  &0.79 &0.76  && 0.77&0.74 &  &0.63 & 0.62 \\
LSTM  &  0.79& 0.76 & & 0.73& 0.68 &&0.74& 0.72 & &0.72 &0.68 & &0.42&0.39\\
\hline
\end{tabular}
}
\label{tabPP}
\end{table}

As the next step we combine all the features of the four cities to predict PM$_{2.5}$ levels (see columns from 3 to 10 in Table~\ref{tabPP}). (Note that we also add four new additional features to keep track of city information in the models. For example, \{1, 0, 0, 0\} means that the data belongs to Beijing.) Although, combination of all the features of the four cities increases the size of the data by a multiple of four, it also increases the number of rows which have a missing value or values in its features or levels tremendously. We could able to get a similar classification accuracy with single city case using NN (see ``All features" column). However, LR and LSTM performed worse as LR is not complex enough for this task and lengths of sequences which are not interrupted by missing values for LSTM were tremendously decreased by too many missing values. For increasing the accuracy of LSTM we might remove wind speed and direction features of all the cities or just include Beijing's wind speed and wind directions. The performance of LSTM is still not as good as NN's (see ``Best" column of Table~\ref{tabPP}).  One other idea could be to make use of rows with missing values by equating missing values to zero (after the normalization) for the features and creating a new feature and a class  for missing PM$_{2.5}$ levels (for example if PM$_{2.5}$ level is missing then the new feature and class will be 0 and 7). The classification accuracies are quite similar to ``All features" case (see ``NA Incl." column of Table~\ref{tabPP}). Finally, we changed the criteria from maximizing the likelihood to mean squared error (MSE) in the machine learning methods. The classification accuracies reduced dramatically for all the methods (see ``MSE Crit." column of Table~\ref{tabPP}).

\section{Discussion}

\subsection{City selection}
We have selected five cities for demonstrating the methodology to simulate the overall PM$_{2.5}$ pollution risk. However, these five cities were not always in an immediately continuous geographic domain, although they are indeed very close to each other. Our results have shown that there is still a statistical correlation between the PM$_{2.5}$ concentrations. In addition, since the main purpose of this modelling is just to demonstrate a methodology to measure clean-air-at-risk of a portfolio of cities that induced by the PM$_{2.5}$ pollution, we feel this selection will still facilitate this purpose. For the same reason, we selected only four cities for demonstrating how to implement a state-of-the-art machine learning technique (LSTM) which employs the dependence of current pollution to the features of current and previous time points.

\subsection{Uncertainty}
The fitted dependence structure for PM$_{2.5}$ concentrations and the marginal distribution parameters are for the dataset we collected. For different cities, or at different times, the best fitting copula and its parameters along with marginal distribution parameters might change. For the prediction section, although we expect LSTM should have the best classification accuracies, the small dataset and abundance of missing values have lowered the performance of deep learning. Therefore, if we have a much larger dataset, the advantage of LSTM should appear.

\subsection{Implications}
This study have profound implications on environmental planning and related policy making. The clean-air-at-risk analysis and its efficient simulation can help evaluate the on-going or potential risk of the PM$_{2.5}$ and the methodology on the implementation of deep learning (LSTM) for PM$_{2.5}$ level classification can facilitate the evaluation of potential pollution and contribute to the selection of mitigating measures.

\section{Conclusion}
In the first part of this paper we introduced an efficient simulation model for quantifying the risk measures of pollution risk in the presence of dependence of PM$_{2.5}$ concentration of cities. The model is based on a copula dependence structure. For assessing model parameters, we analyzed a limited data set of PM$_{2.5}$ levels of Beijing, Tianjin, Chengde, Hengshui, and Xingtai. This process revealed a better fit for the t-copula dependence structure with generalized hyperbolic marginal distributions for the PM$_{2.5}$ log-ratios of the cities. Furthermore, we adopted the efficient simulation strategies, IS and SIS, developed for financial risk management to compute CaR and CCaR under the t-copula dependence structure. Our numerical results showed that the proposed methods are much more efficient than a naive simulation for computing the exceeding probabilities and conditional excesses.

In the second part, we implemented three machine learning methods to predict PM$_{2.5}$ levels of the next three-hour period of four Chinese cities, Beijing, Chengde, Xingtai, and Zhangjiakou. For this purpose, we used the pollution and weather data collected from the stations located in these four cities. Instead of coding the machine learning algorithms, we employed a state-of-the-art machine learning library, Torch7. This allowed us to try out one of the best performing machine learning methods of fields like speech recognition of the current time, namely LSTM. Unfortunately, due to small data size and lots of missing values (when we combined the features of the cities) in the data, LSTM did not perform better than a multilayer perceptron.  However, we still were able to get a classification accuracy above 0.72 on the test data. 

\section*{Acknowledgements}
This work was supported by Xi'an Jiaotong-Liverpool University Research Fund Project RDF-14-01-33.

\end{document}